\documentclass[journal]{IEEEtran}

\usepackage{cite}
\usepackage{array}
\usepackage{color}
\usepackage{float}
\usepackage[cmex10]{amsmath}
\usepackage{nomencl}						
\usepackage[normalem]{ulem}			
\usepackage{diagbox}						
\usepackage{slashbox}						
\usepackage{colortbl}						
\usepackage{multirow}						
\usepackage{tabularx}
\usepackage{placeins}
\usepackage{algorithm}
\usepackage{mathrsfs}
\usepackage{mathdots}

\usepackage{amssymb}
\usepackage{amsthm}
\usepackage{arydshln}
\usepackage{color}
\usepackage[noend]{algpseudocode}
\usepackage{bm}
\usepackage{url}
\usepackage[hidelinks]{hyperref}
\usepackage[T1]{fontenc}
\newcommand{\subparagraph}{}
\usepackage{fancyhdr} 
\usepackage{extarrows}
\usepackage[switch,columnwise]{lineno}
\usepackage{MnSymbol}
\usepackage[noend]{algpseudocode}
\usepackage{siunitx}
\usepackage{cleveref}

\usepackage{gensymb}

\usepackage{array, makecell}

\ifCLASSINFOpdf
  \usepackage[pdftex]{graphicx}
	\graphicspath{{./figure/}}
  \DeclareGraphicsExtensions{.pdf,.jpeg,.png}
\else
  \usepackage[dvips]{graphicx}
  \graphicspath{{./figure/}}
  \DeclareGraphicsExtensions{.eps}
\fi

\ifCLASSOPTIONcompsoc
  \usepackage[caption=false,font=normalsize,labelfont=sf,textfont=sf]{subfig}
\else
  \usepackage[caption=false,font=footnotesize]{subfig}
\fi

\newcommand{\figref}[1]{\textcolor{black}{Fig.~\ref{#1}}}

\newcommand{\equref}[1]{\textcolor{black}{(\ref{#1})}}

\newcommand{\appendixref}[1]{\textcolor{black}{Appendix~\ref{#1}}}

\begin{document}
\bstctlcite{IEEEexample:BSTcontrol}
\setcounter{page}{1}

\title{The Intrinsic Communication in Power Systems: A New Perspective to Understand Synchronization Stability}
\author{Yitong Li, \IEEEmembership{Member, IEEE}, Timothy C. Green, \IEEEmembership{Fellow, IEEE}, Yunjie Gu, \IEEEmembership{Senior Member, IEEE}}

\ifCLASSOPTIONpeerreview
	\maketitle 
\else
	\maketitle
\fi


\begin{abstract}
Synchronization is an essential element in three-phase ac electric power systems. The large-scale integration of converter-interfaced resources leads to the power grid transformation from voltage-source-dominated to voltage-current-source-composite, which also raises new challenges to model and analyze the system synchronization. In this article, we present the intrinsic analogy of a power system to a communication system, which is here called power-communication isomorphism. Based on this isomorphism, we revisit power system synchronization stability from a communication perspective and thereby establish a theory that unifies the synchronization dynamics of heterogeneous power apparatuses. In particular, the proposed theory is used to interpret and analyze, for example, the different power control speeds of power apparatuses; the role of network line dynamics in synchronization analysis; and the synchronization capability of current-source-dominated (e.g., grid-following-converter-dominated) grids. The findings are verified on the IEEE 68-bus test system.
\end{abstract}




\section{Introduction} \label{Section:Introduction}

Driven by the imperative of decarbonisation and clean growth, the primary energy of electric power systems is transforming from fossil fuels to renewable resources. The change of the primary energy is accompanied by a change of technologies for power generation and conversion. Renewable resources, mainly wind and solar energy, as well as grid-scale battery storage plants, are interfaced to power systems by power electronic converters instead of conventional synchronous generators. The increasing penetration of converter-interfaced resources poses new threats to system stability. Converter-induced oscillations have been reported worldwide, many of which had major consequences. For example, the 2019 power outage in UK was, in part, triggered by a sub-synchronous oscillation of wind turbine converters in Hornsea windfarm according to the report provided by National Grid UK \cite{bialek2020does}. The underpinning mechanisms of power system stability are not fully understood and this has drawn international attention.

The stability of a power system is defined as the ability to keep all apparatus in the system synchronized to a single frequency with power flows and voltage profiles throughout the system within some expected range \cite{kundur1994power,kundur2004definition}. The classic synchronization stability theory for power systems is tailor-made for synchronous generators which are govern by the physical law of the motion of their rotors. However, converters are governed by control algorithms, which gives increased flexibility, and therefore complexity, in converter behaviours \cite{rocabert2012control}. Up to now, the control-defined behaviour of converters are categorised into two classes. The first class, called grid-forming \cite{rosso2021grid,rocabert2012control,matevosyan2019grid,li2022revisiting}, behaves as a voltage source which synchronizes to the grid according to power balancing. The second class, called grid-following \cite{rocabert2012control,li2022revisiting,fan2020identifying,wen2015analysis}, behaves as a current source (or sink) which synchronizes to the grid according to voltage signals. This forming-following dichotomy creates a heterogeneous grid that sets a barrier for whole-system synchronization stability analysis \cite{wang2020grid,sof2018system}.

To address the aforementioned challenges, we looked again at the nature of synchronization in three-phase ac electric power systems. We illustrate that there is an intrinsic communication mechanism underlying power systems, which is described as a \textit{power-communication isomorphism}. Based on the isomorphism, we developed a theory that creates new insights into power system synchronization dynamics. Based on it, we can intuitively understand the different power control speeds of voltage- and current-type power apparatuses; the role of electromagnetic transient (EMT) dynamics of power network lines; the synchronization stability of a 100\%-grid-following-converter grid; and how the system stability changes if adding grid-forming apparatuses into grid-following-converter-dominated grids.

\section{Power-Communication Isomorphism} \label{Section:PowerCommunication}

\begin{figure*}
    \centering
	\includegraphics[scale=0.9]{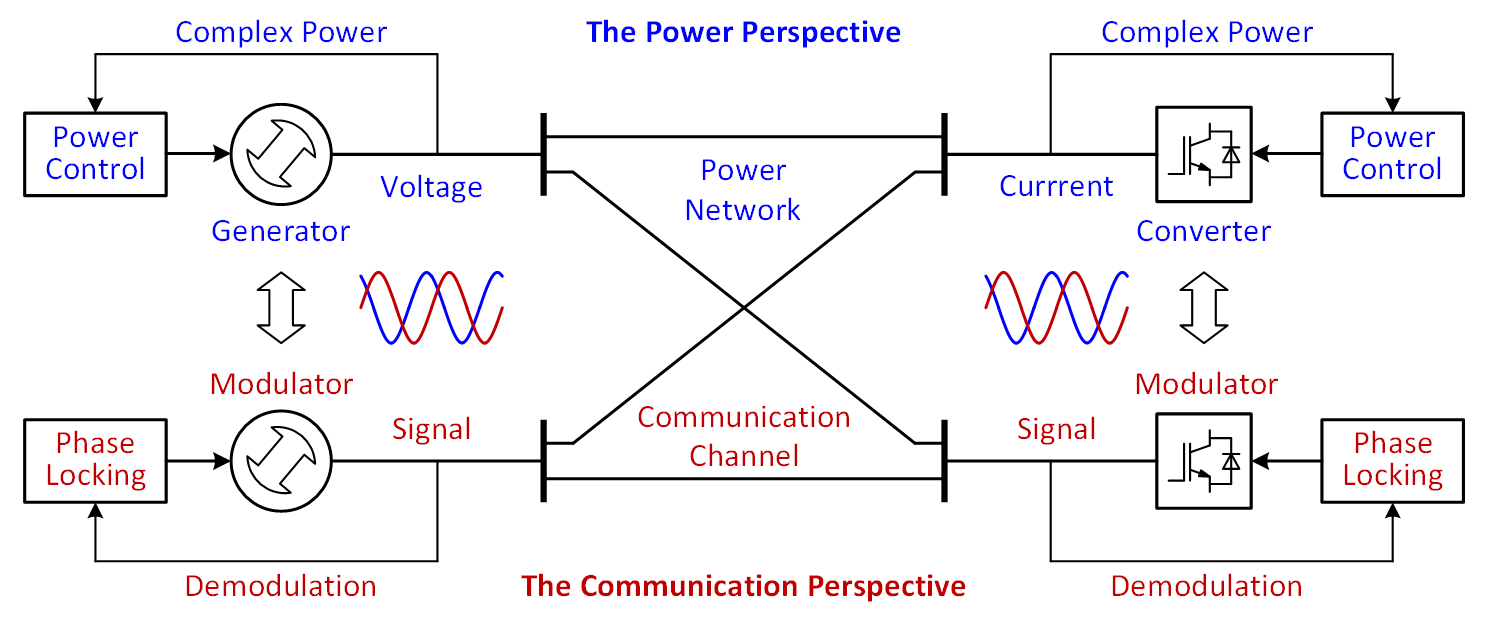}
	\caption{Illustration of power-communication isomorphism in power systems. The upper part shows a part of system viewed from the perspective of power transfer and the lower part is viewed from the perspective of communication.}
	\label{fig_iso}
\end{figure*}

The concept of power-communication isomorphism is illustrated in \figref{fig_iso}. The voltages and currents in a power system are viewed as communication signals carrying both energy and information. The power apparatuses, including generators and converters, serve as modulators in that they create three-phase sinusoidal signals from internal oscillators (rotors, controllers, etc). The amplitude, frequency and phase of an internal oscillator are base-band signals which are shifted to the carrier-band of 50 Hz or 60 Hz via frame transformation or rotation, creating an effect equivalent to amplitude and angle modulation. Mathematically, a three-phase signal is represented as a complex number $A e^{j \theta}$, where $A$ and $\theta$ are the amplitude and angle of the signal \cite{harnefors2007modeling,briz2000analysis,martin2004complex,milano2021complex}. The amplitude and angle can be combined into a complex phase defined as $\vartheta = \ln A + j \theta$, so the amplitude-angle modulation is jointly expressed as a complex exponential function
\begin{equation}
e^\vartheta = A e^{j \theta} = A (\sin{\theta } + j \cos{\theta})
\end{equation}
The time-derivative of the $\vartheta$ is called the complex frequency $\varpi = \dot{\vartheta} = A^{-1}\dot{A} + j \omega$, whose real part reflects amplitude variation, and the imaginary part $\omega=\dot{\theta}$ is the angular frequency.

The modulated signals are propagated over a power network and that network can be viewed as a set of communication channels. The channels include all passive components in the power network, including transmission lines, transformers, series/shunt compensators, harmonic filters, and passive loads. The inner control loops of converters can be represented as equivalent impedances in series or shunt with the sources, and therefore can also be counted among the channels. The active apparatuses, including generators and converters, are defined as nodes that interact (communicate) over the channels. There are two types of nodes in a power system. A voltage node applies a voltage source to the network, and represents grid-forming apparatuses, including synchronous generators and grid-forming converters \cite{li2022revisiting}. A current node applies a current source to the network, and represents grid-following converters \cite{rocabert2012control,pattabiraman2018comparison,li2022revisiting}. From the communication point of view, a voltage node transmits a voltage signal to the network and receives a current signal, and a current node does the opposite. There may be multiple nodes in the network and their signals are received at each node as a single signal by superposition. The complex power seen at a node is defined as \cite{czarnecki2004some}
\begin{equation}        \label{eq_S}
   S = e^{\vartheta_\text{rx}} e^{{{\vartheta_\text{tx}}^{\!\!\!*}}} = A_\text{tx} A_\text{rx} \, e^{j(\theta_\text{rx} - \theta_\text{tx})}
\end{equation}
where $e^{{{\vartheta_\text{tx}}}}$ is the transmitted signal, $e^{\vartheta_\text{rx}}$ is the received signal, and the superscript $^*$ denotes complex conjugation. It is worth noting that, $A_\text{rx}e^{\theta_\text{rx}}$ and $A_\text{tx}e^{\theta_\text{tx}}$ are ac signals, but $S$ can be dc at steady-state if $\omega_\text{rx} = \dot{\theta}_\text{rx}$ equals to $\omega_\text{tx} = \dot{\theta}_\text{tx}$ and $\theta_\text{rx}-\theta_\text{tx}$ is constant. In other words, the complex power has a demodulation effect that converts a carrier-band signal to base-band. 

\section{Power Apparatuses and Modulator Dynamics}

Synchronization is an essential element for all power apparatuses in ac power grids. There are heterogeneous synchronization schemes co-existing in power systems which present an obstacle to systemic analysis. Voltage nodes use power-based synchronization via physical or emulated rotors \cite{zhang2009power,2014virtual,li2022revisiting}, whereas current nodes use signal-based synchronization via phase-locked loops (PLLs) \cite{chung2000phase,li2022revisiting}. The theory of power-communication isomorphism illuminates the unified synchronization principle underlying the power- and signal-based synchronization schemes. 

\begin{figure}
    \centering
	\includegraphics[width=0.37\textwidth]{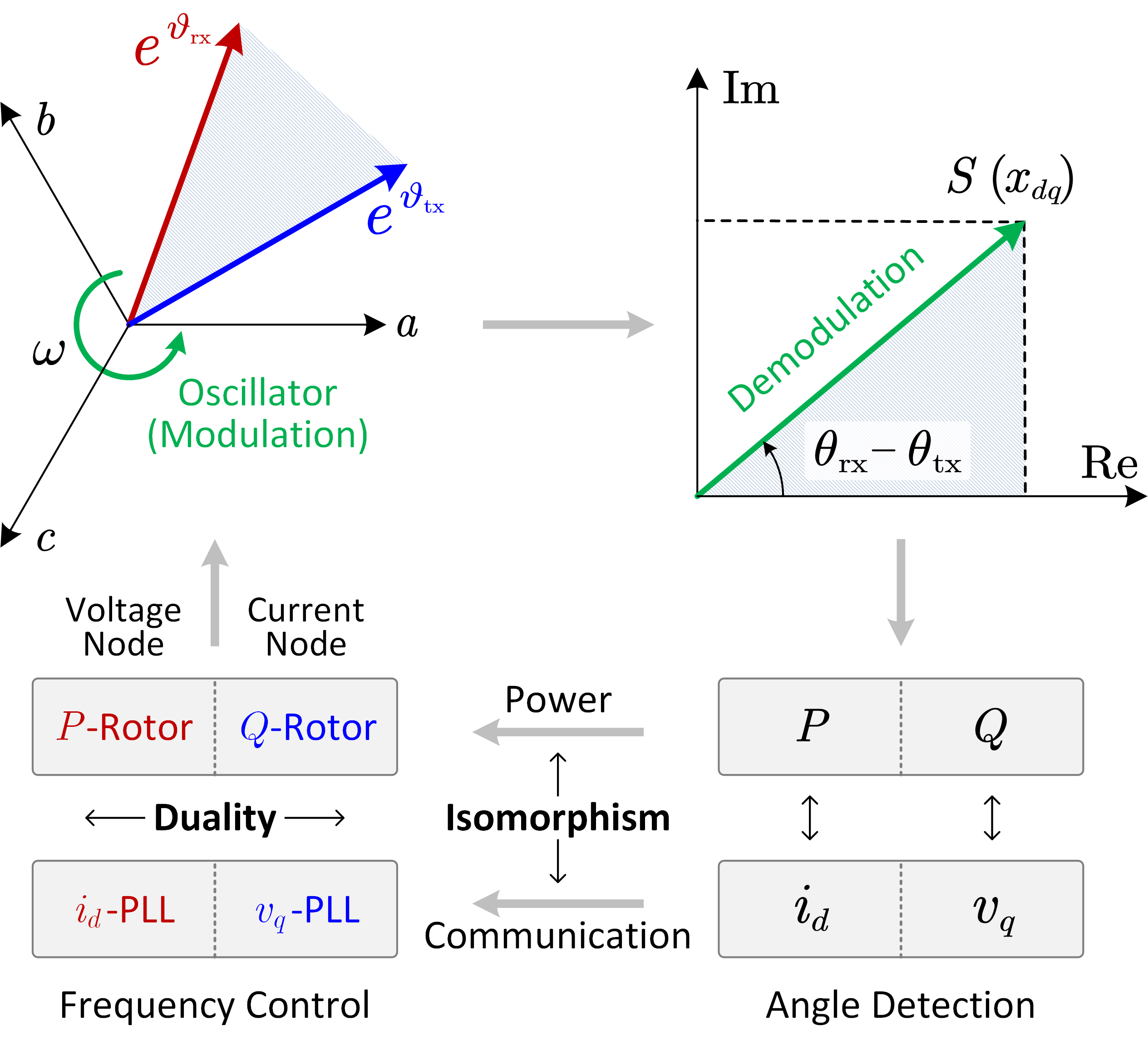}
	\caption{Unified synchronization principle in the light of power-communication isomorphism: the equivalence of rotor and PLL, and the duality of voltage and current nodes. We use a prefix to distinguish the conventional rotor and PLL ($P$-Rotor, $v_q$-PLL) and the ones derived from the isomorphism ($Q$-Rotor, $i_d$-PLL).}
	\label{fig_pll}
\end{figure}

The essence of synchronization is to detect the angle difference between nodes and mitigate the difference via feedback control of the frequency of the oscillators of the nodes in question. In the light of power-communication isomorphism, the demodulated complex power $S$ is a natural candidate for angle detectors. However, $S$ is a complex number, so its real and imaginary parts are used in practice. For a voltage node, the active power $P=\text{Re}(S)$ is used as an angle detector; for a current node, the $q$-axis voltage $v_q=\text{Im}(v_{dq})$ is used as an angle detector. It is worth noticing that $P = i_d A_v \propto i_d$ and $Q = v_q A_i \propto v_q$, which implies the power calculation is also equivant to the $dq$-frame Park transformation, and further implies an unified synchronization principle illustrated in \figref{fig_pll}. The rotor of a voltage node is equivalent to a PLL synchronising to the current $i_d$, and the PLL of a current node is equivalent to a rotor accelerating/decelerating under reactive power $Q$. This gives the duality relationship between the synchronization of voltage and current nodes \cite{li2022revisiting}.

Angle detection via $P$ and $Q$ can be generalized by introducing a projection angle $\mu$
\begin{equation}    \label{eq_W}
W = \mathrm{Re}( e^{-j\mu} \, S)
\end{equation}
where $W$ is named generalized power, and the associated frequency control is governed by a generalized rotor. If $\mu$ is set to $0$ or $\pi/2$, $W$ reduces to $P$ or $Q$ respectively. Then the frequency $\omega$ of an apparatus can be given by $W$ through the apparatus synchronization dynamics (e.g., physical rotor inertia and damping of synchronous generators or virtual inertia and damping of converters \cite{2014virtual,li2022revisiting}) as, 
\begin{equation}   \label{eq_SwingEqu}
H\dot{\omega} = W^*-W-D\omega
\end{equation}
where $H$ is inertia, $D$ is damping, $W^*$ is generalized power reference. This unifies the synchronization dynamics of all apparatuses in power systems, or equivalently the modulation dynamics of all modulators from the communication perspective.

\section{Power Networks and Channel Dynamics}

\begin{figure*}
    \centering
	\includegraphics[scale=0.9]{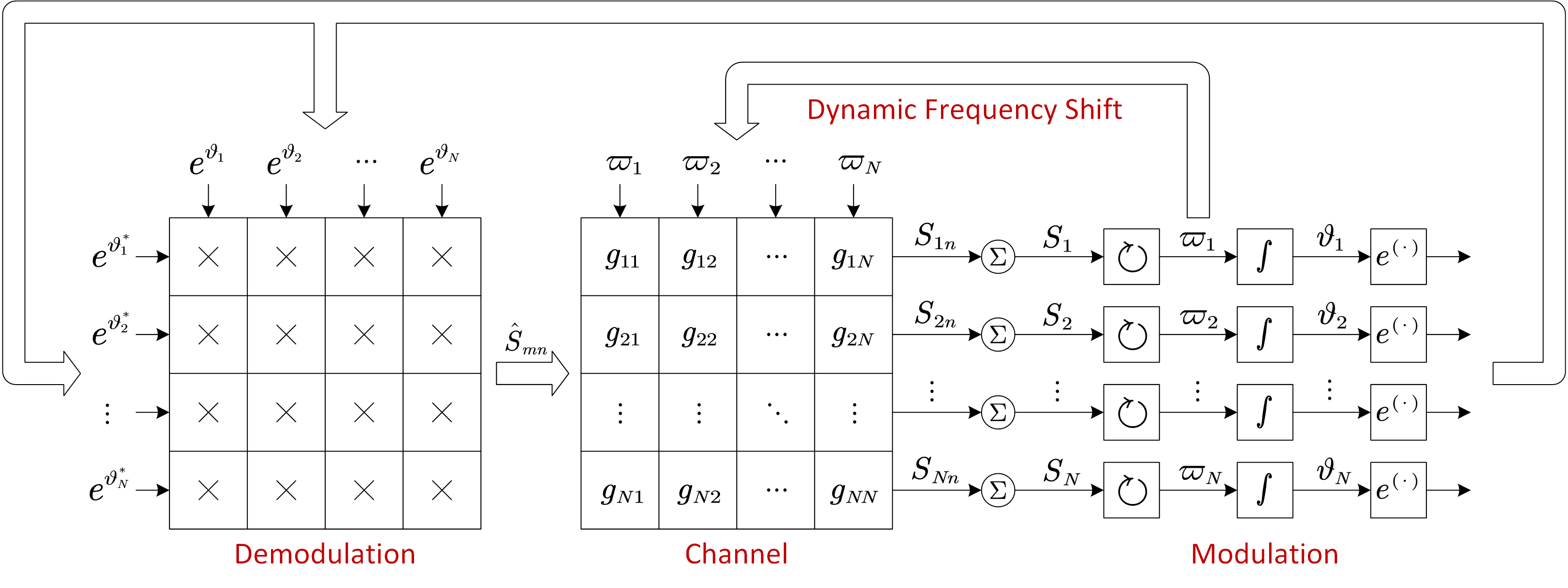}
	\caption{The overall model of the power-communication isomorphic system considering the dynamic channel gains. Symbols: $*$ denotes complex conjugation, $\times$ denotes multiplication, $\sum$ denotes summation, $\int$ denotes integration, and $\circlearrowright$ denotes the oscillator that synthesise the internal complex frequency according to the balancing of the complex power.}
	\label{fig_ch}
\end{figure*}

In a communication system, the dynamics of a communication channel determine the maximum communication rates according to the Shannon-Hartley theorem \cite{chen2004electrical}. In this section, we investigate the role of channel dynamics in power systems.

\subsection{Power Network and Communication Channel}

We use $g_{mn}$ to represent the channel gain when a signal (current or voltage) is transmitted from node $n$ to node $m$. By varying $m$ and $n$, we can get a channel gain matrix $[g]$, which reflects the communication topology of the network. On the other hand, the dynamic nodal admittance matrix $[Y]$ defines the power (physical) topology of the network \cite{kundur1994power,gu2020impedance}, where the "dynamic" means $[Y]$ is a transfer function matrix \cite{gu2020impedance} rather than a complex-value nodal admittance matrix conventionally used in power flow analysis \cite{kundur1994power}. We show how the communication and power topology are related next. 

Suppose that the voltage nodes in the network are numbered by $\{1,2,\cdots,N_v\}$ and current nodes by $\{N_v+1,N_v+2,\cdots,N_v+N_i\}$, where $N_v$ and $N_i$ are the total number of voltage and current nodes respectively. We partition $[Y]$ at the $N_v$-th row and $N_v$-th column:
\begin{equation}
    [Y] = 
    \begin{bmatrix}
        [Y_{vv}] & [Y_{vi}] \\
        [Y_{iv}] & [Y_{ii}] \\
    \end{bmatrix}
\end{equation}
and we have
\begin{equation}
\label{eq_Yiv}
\begin{array}{ll}
    [i_v] = [Y_{vv}][v_v] + [Y_{vi}][v_i]  \\ \relax
    [i_i] = [Y_{iv}][v_v] + [Y_{ii}][v_i]
\end{array}
\end{equation}
where $[v_v]$ and $[v_i]$ are vectors representing the voltages at the voltage and current nodes respectively, and $[i_v]$ and $[i_i]$ are the corresponding current vectors. From the communication perspective, $[v_v]$ and $[i_i]$ are transmitted by nodes, and $[i_v]$ and $[v_i]$ are received from the network. Thus we rearrange (\ref{eq_Yiv}) to show the mapping from $[v_v,i_i]$ to $[i_v,v_i]$
\begin{equation}
\label{eq_Hiv}
\begin{array}{ll}
    [i_v] = \left([Y_{vv}] - [Y_{vi}] [Y_{ii}]^{-1} [Y_{iv}]\right)[v_v] + [Y_{vi}][Y_{ii}]^{-1}[i_i]  \\ \relax
    [v_i] = -[Y_{ii}]^{-1} [Y_{iv}][v_v] + [Y_{ii}]^{-1}[i_i]
\end{array}
\end{equation}
from which follows 
\begin{equation}
\label{eq_giv}
    [g] = 
    \begin{bmatrix}
        [Y_{vv}] - [Y_{vi}] [Y_{ii}]^{-1} [Y_{iv}] & [Y_{vi}][Y_{ii}]^{-1} \\
        -[Y_{ii}]^{-1} [Y_{iv}] & [Y_{ii}]^{-1} \\
    \end{bmatrix}.
\end{equation}
i.e., the channel gain matrix. If the network only contains voltage nodes, $[g] = [Y]$, indicating that the communication and power topology are identical. Otherwise, the power topology $[Y]$ is twisted in the communication topology $[g]$ due to the interaction of current and voltage nodes.

\subsection{Dynamic Frequency Shift}

A channel (e.g., a transmission line) consists of linear circuits and therefore can be represented as a linear transfer function $G(s)$ in frequency domain. Since the modulation-demodulation has a frequency shifting effect, the equivalent transfer function seen by the base-band signals is shifted to $G(j \omega_\text{c} + s)$ according to Fourier analysis \cite{oppeitheim1983signals}. This frequency shift representation assumes a constant carrier frequency which does not hold for a power system where different nodes may have different instantaneous carrier frequencies that are varying in real-time subject to load balancing. To address this issue, we propose a new time-domain representation for a channel. The time-domain gain of a channel is defined as $g = e^{{\vartheta ^\prime}}/e^{{\vartheta}}$ where $e^{{\vartheta}}$ and $e^{{\vartheta ^\prime}}$ are the input and output signals of the channel respectively. The alternating part of $e^{{\vartheta}}$ and $e^{{\vartheta ^\prime}}$ are cancelled in the division so $g$ is a base-band variable. $G(s)$ can be decomposed into a series of first-order systems $G(s) = \sum_k (s - p_k)^{-1} a_k$, each of which induces a gain $g_k$, and the total gain is the sum of all $g_k$. We simply investigate one of the first-order systems $G(s) = (s - p)^{-1} a$ without losing generality, where $p$ is the pole of $G(s)$ and $a$ is the coefficient. The corresponding differential equation for the signal passing the channel is
\begin{equation}
\label{eq_diff}
d \, e^{{\vartheta ^\prime}}/dt = p \, e^{{\vartheta ^\prime}} + a \, e^{{\vartheta}}
\end{equation}
which yields the differential equation for the channel gain $g$ 
\begin{equation}
\label{eq_gain_diff}
d {g}/dt = (p - \varpi) g + a.
\end{equation}
It is clear that $g$ depends on $\varpi$ dynamically, and this effect is named \textit{dynamic frequency shift} as an extension to the (static) frequency shift in Fourier analysis. 

\subsection{Whole-System Model}

Taking into account the dynamic channel gain, the overall model for the power-communication isomorphic system is illustrated in \figref{fig_ch}. For a network with $N$ nodes, the $n$-th node transmits a signal $e^{\vartheta_n}$ to the network which is demodulated by another node $m$ to yield $\hat S_{mn} = e^{{\vartheta_n}} e^{{\vartheta_m} ^{\!\!\!\!\!*}}$. We put $\,\hat{}\,$ on $\hat{S}_{mn}$ because it is not physical power as the channel gain is not yet included. Considering the channel gain, the complex power transfer over the channel from the $n$-th node to the $m$-th node is ${S}_{mn}=g_{mn} \hat{S}_{mn}$ where $g_{mn}$ is the corresponding channel gain. All traffic in the network shares the channels according to the superposition principle, so the total complex power received at node $m$ is the summation: $S_{m} = \sum_n  S_{mn}$ \footnote{For example, for a two-bus system, the total complex power at bus 1 is $S_{1}=S_{11}+S_{12}$. Just be careful that $S_{11}$ should also be considered to get the total complex power, based on the superposition theorem.}. The total complex power $S_{m}$ is fed to the oscillator of the $m$-th node. The oscillator governs the complex frequency $\varpi_{m}$ which is integrated to the complex angle $\vartheta_{m}$. The signal $e^{\vartheta_m}$ is modulated from $\vartheta_{m}$ and transmitted to channels, and thus closes the loop of modulation-demodulation.

\subsection{Channel Bandwidth}

The dynamic frequency shift illuminates interesting properties of base-band signal propagation over a channel. To illustrate this, we find the linearized solution of $g$ from (\ref{eq_gain_diff}) and put it into $S_{mn}$
\begin{equation}    \label{eq_pert}
   \Delta S_{mn} = S_{mn0}(\Delta {\vartheta_m}^{\!\!\!\!\!*} + F \cdot \Delta {\vartheta_n})
\end{equation}
where the prefix $\Delta$ and subscript $0$ denote the perturbation and equilibrium of a dynamic variable, and $F$ is a low pass filter
\begin{equation}    \label{eq_filter}
F(s) = \frac{ j\omega_0 - p}{s + j\omega_0 - p}.
\end{equation}
The detailed derivation of \equref{eq_pert} and \equref{eq_filter} can be found in \appendixref{Appendix:ChannelBandwidth}. It is clear from (\ref{eq_pert}) that a channel induces asymmetry in base-band signal propagation. The angle perturbation $\Delta \vartheta_m$ at the receiving end $m$ affects the complex power $\Delta  S_{mn}$ instantaneously, whereas the angle perturbation $\Delta \vartheta_n$ at the transmitting end $n$ passes through a low-pass filter $F$ before affecting $\Delta  S_{mn}$ \footnote{Just be careful that $\Delta S_{mn}$ with $n=m$ also satisfies \equref{eq_pert}, i.e., $\Delta S_{mm} = S_{mm0}(\Delta {\vartheta_m}^{\!\!\!\!\!*} + F \cdot \Delta {\vartheta_m})$. This gives the self-complex power at bus $m$.}. Thus we defined the \textit{channel bandwidth} as 
\begin{equation}
\begin{array}{c}
\omega_\text{b} = \text{sup} \, |\omega|, \,\, \text{subject to} \\
|F(j\omega)| > 1/\sqrt{2} \,\,\, \text{and} \,\,\, |\angle F(j\omega)| < \pi/4. 
\end{array}
\end{equation}

The channel bandwidth identifies the limit speed of power transfer and angle synchronization on the channel. Within the channel bandwidth, the dynamic channel gain $g$ responds almost instantaneously to $\varpi$. In such a case, $g$ can be approximated by letting $dg/dt = 0$ in (\ref{eq_gain_diff}) which yields
\begin{equation}    \label{eq_ChanneldynamicApprox}
    g \approx (\varpi - p)^{-1} a = G(\varpi).
\end{equation}

The base-band signals are equivalent to phasors, and the channel dynamics are equivalent to EMTs in power systems. For example, an inductive line is modelled as $j\omega_0 L=jX$ in phasor-domain analysis, where $X$ is a steady-state reactance, i.e., a constant value. By contrast, the same inductive line is modelled as $j\omega L$ where $\omega$ is variable in EMT analysis \cite{kundur1994power}. In other words, if we set $g_{mn} = G(\varpi_0)$ which gives phasor-domain steady-state channel network, and if we set $g_{mn} = G(\varpi_n)$ which includes the dynamics from $\varpi_n$ to channel and therefore the so-called EMT channel dynamics.

\section{Rethinking Synchronization Stability of Power Systems} \label{Section:PowerSystemStability}

The power-communication isomorphism theory provides a new perspective to rethink power system synchronization stability, as elaborated next.

\begin{figure*}
    \centering
	\includegraphics[scale=0.60]{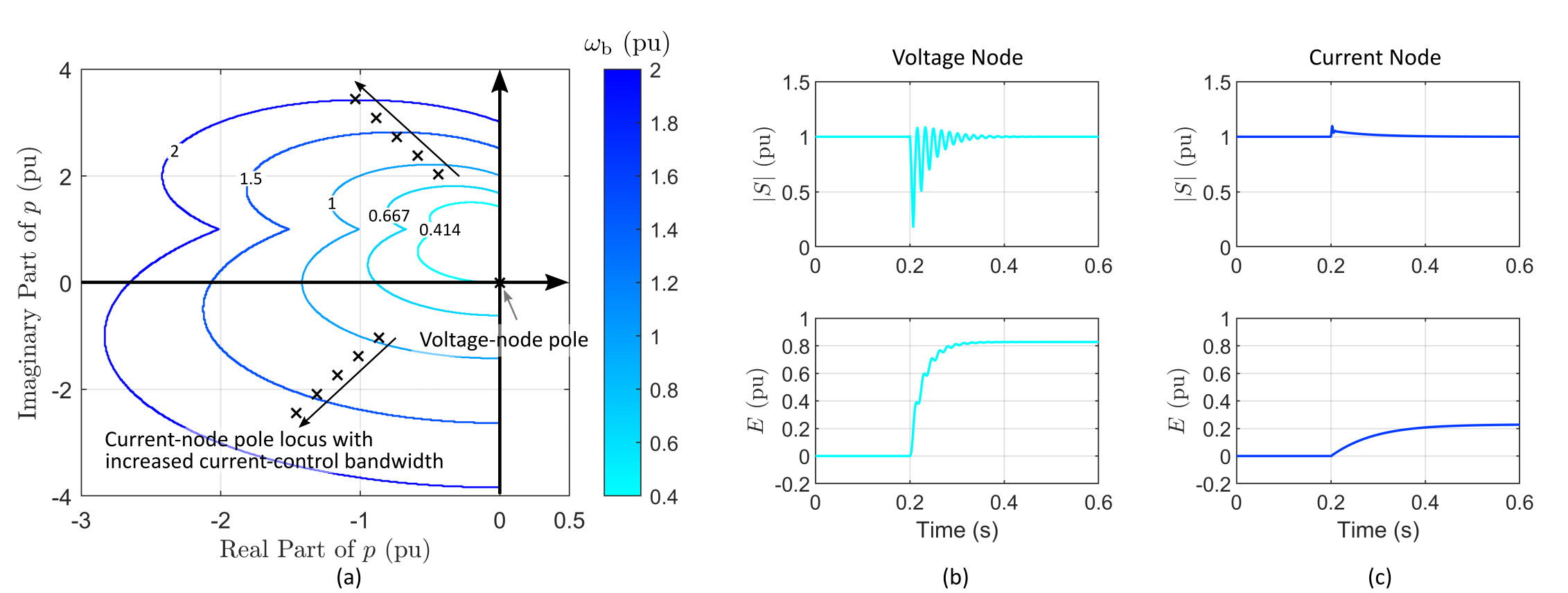}
	\caption{Channel bandwidth and its impact on transient power. (a) Equi-bandwidth contours on the plane of channel pole indicating that current nodes have higher channel bandwidth $\omega_\text{b}$ than voltage nodes. (b)-(c) Transient power ($|S|$) and the accumulated energy ($E$) at a node subject to a phase jump at the remote end: (b) for a voltage node and (c) for a current node. The relatively low channel bandwidth associated with a voltage node results in high accumulated energy and requires a large energy cache. Variables in the figure are presented per unit.}
	\label{fig_bandwidth}
\end{figure*}

\subsection{Power Control Speed and Channel Bandwidth}

The channel bandwidth determines the maximum speed of power transfer and angle synchronization over the network. The channel bandwidth $\omega_\text{b}$ is determined by the pole of the channel. There are four types of channels in a network, namely voltage-voltage, voltage-current, current-voltage, current-current channels. A voltage-voltage channel is the channel between two voltage nodes, and other types are defined similarly. A voltage-voltage channel is generally an inductive transmission line with a very small resistance, so its pole is approximately zero and the corresponding channel bandwidth is $0.41\omega_0$ ($\omega_0$ is 50 or 60 Hz). Other channels where current nodes are associated (i.e., voltage-current, current-voltage, and current-current channels) have a negative-real pole so their channel bandwidths are higher, as marked in \figref{fig_bandwidth} (a). This is because the parallel connected passive loads and the current control loops induce equivalent shunt resistances at the current node which damps the channel pole. Due to the different channel bandwidths, a voltage node normally has a low power control speed (e.g., large inertia of synchronous generators) and current node normally has a high power control speed (e.g., fast current and power control of grid-following inverters) in practice. This also results in high transient power and energy perturbation of voltage node [see \figref{fig_bandwidth} (b)], and low transient power and energy perturbation of current node [see \figref{fig_bandwidth} (c)].

\begin{figure}[t!]
    \centering
	\includegraphics[scale=0.6]{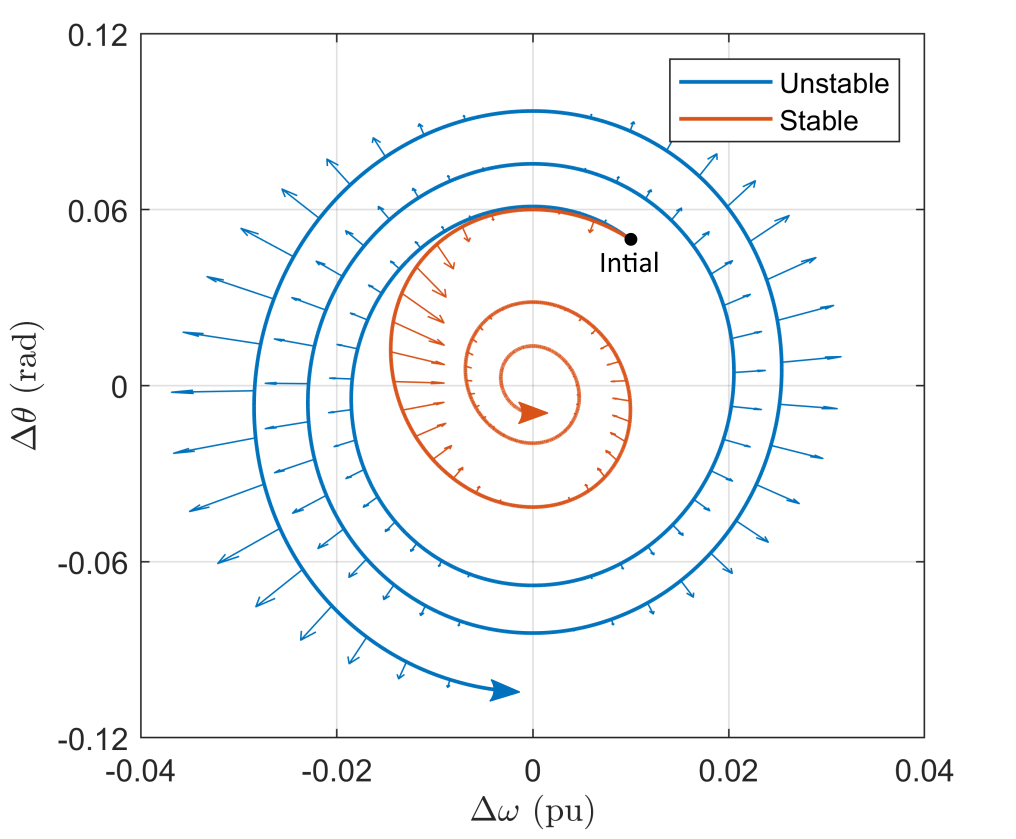}
	\caption{Dynamic frequency shift may induce negative damping. The negative $\partial W / \partial \omega$ results in a vector field on the $\omega\text{-}\theta$ phase plane pointing outward (shown as the outward arrow on the blue trajectory), which causes the trajectory to diverge from the equilibrium and makes the system unstable. This problem can be solved by injecting extra positive damping in frequency control, to make the vector field pointing inward (shown as the inward arrow on the amber trajectory) .}
	\label{fig_gamma_IBR}
\end{figure}

\begin{figure*}[t!]
    \centering
	\includegraphics[scale=0.75]{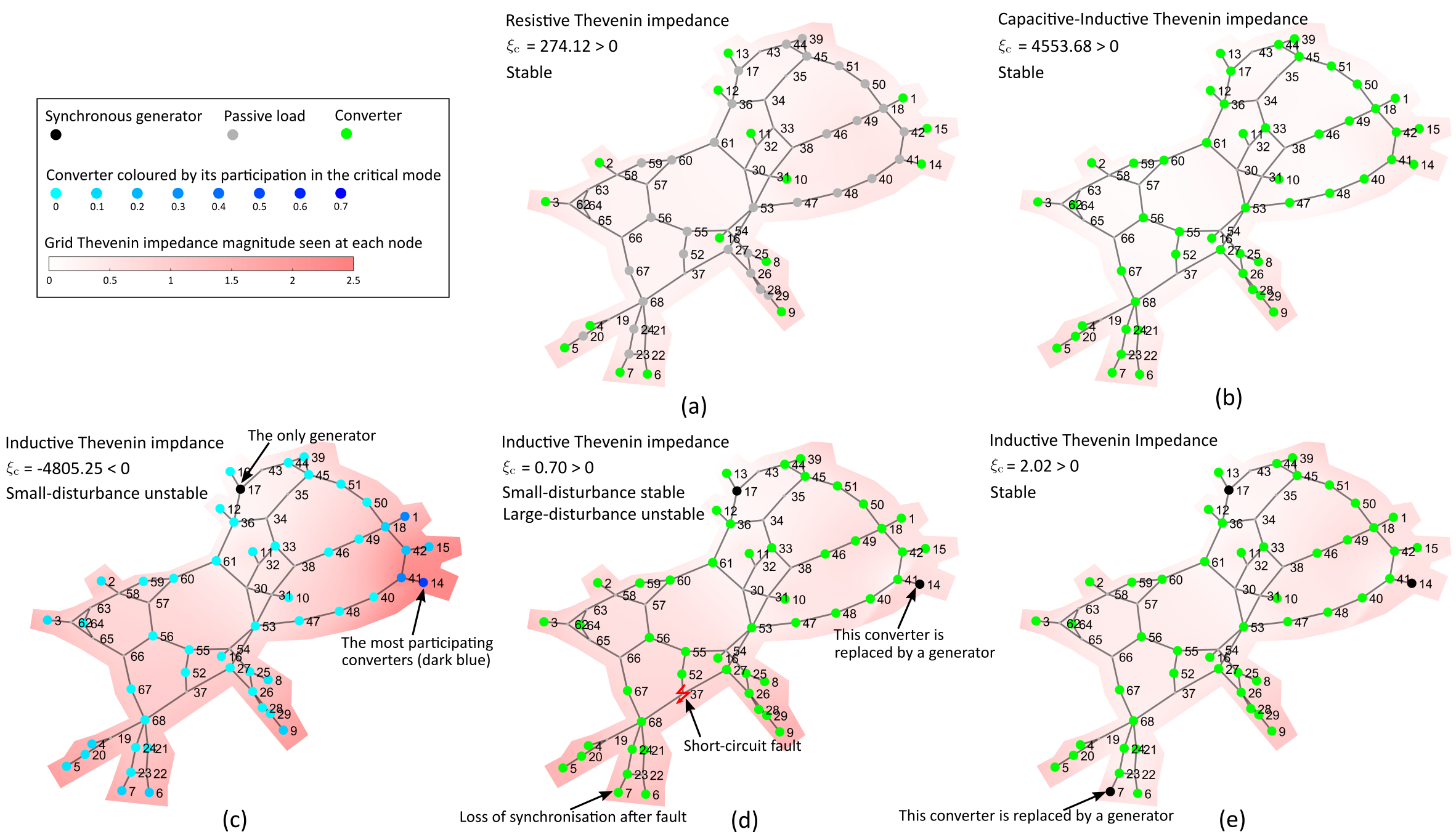}
	\caption{Summary of test results on the IEEE 68-bus system with varying numbers of grid-following converters (green dots), synchronous generators (black dots) and passive loads (grey dots). Heat-map in shades of red indicates impedance magnitude seen at each node. The small-disturbance and large-disturbance stability of the tested system is marked via texts on the map, and the detailed transient responses are shown in \figref{fig_K_SimResult}. (a)  17 converters, 0 synchronous generators, with passive loads. (b) Converters at all 68 busses with active loads. (c) 1 synchronous generator (at node 17). (d) 2 synchronous generators (at nodes 14,17). (e) 3 synchronous generators (at nodes 7,14,17). }
	\label{fig_K}
\end{figure*}

\begin{figure*}[t!]
    \centering
	\includegraphics[scale=0.85]{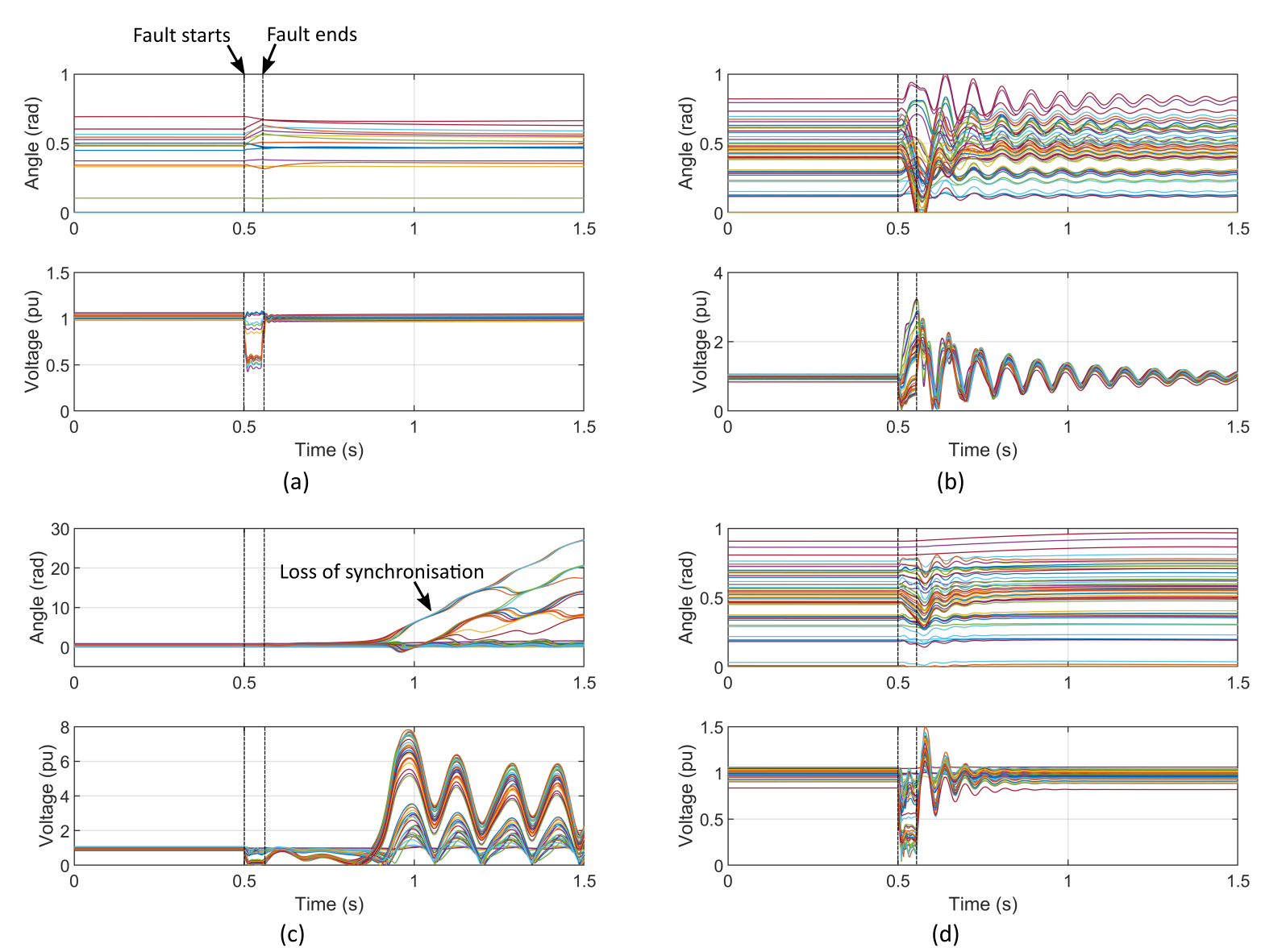}
	\caption{Time-domain EMT simulations of the modified IEEE 68-bus system under a short-circuit fault, which occurs at bus 37 at 0.5 s and is cleared after 3 fundamental cycles, i.e., at 0.55 s. (a) 100\% converters (grid-following) with passive loads. (b) 100\% converters with active loads. (c) Two synchronous generators (at node 14,17). (d) Three synchronous generators (at node 7,14,17). The results in (a)-(d) corresponds to tests (a),(b),(d) and (e) in \figref{fig_K}. The simulation result for test (c) is not stable around the equilibrium and is therefore not presented.}
	\label{fig_K_SimResult}
\end{figure*}

\subsection{Synchronization Damping and Dynamic Frequency Shift}

Combining the synchronization principle in \equref{eq_W} and \equref{eq_SwingEqu}, with the channel model in \equref{eq_S} and \figref{fig_iso}, we get the dynamic model for the $m$th arbitrary apparatus with considering whole-system interactions as
\begin{equation}  \label{eq_sync}
\begin{aligned}
H_m \dot{\omega}_m = W_m^{\star} - W_m  - D_m\omega_m
\end{aligned}
\end{equation}
with
\begin{equation}    \label{eq_Wm}
W_m = \text{Re} \left( e^{-j\mu_m} \sum\nolimits_n { g_{mn} A_m A_n e^{-(\theta_m-\theta_n)}} \right)
\end{equation}
$W_m$ depends on $\theta_m$ and $\theta_n$ which gives the synchronizing coefficient as discussed later in next subsection. Here, according to \equref{eq_ChanneldynamicApprox}, the channel gain $g_{mn} \approx G_{mn}(j\omega_n)$ is a function of $\omega_n$, i.e., the dynamic frequency shift of channels. This means $W_m$ also depends on $\omega_n$. $\partial W_m / \partial \omega_n$ tends to induces negative damping. For example, for a synchronous generator, $\partial W / \partial \omega = \partial P / \partial \omega = \partial(V_1V_2\text{sin}(\Delta \theta)/\omega L)/\partial \omega < 0$. This negative damping is essentially introduced by the frequency-dependent line inductor $\omega L$ in EMT analysis, compared to the phasor analysis with assuming a constant $\omega_0L=X$. This negative damping would lead to system instability if the damping $D_m$ of the apparatus itself is not sufficiently large. \figref{fig_gamma_IBR} shows the test result of a synchronous generator with or without enough $D_m$, which validates the analysis. This channel-induced negative damping illuminates the role of EMTs in phasor-domain analysis, which also agrees the literature observing that the EMT line dynamics tend to worsen the system stability \cite{li2022mapping}.

\subsection{Synchronization Coefficient and Modulation-Demodulation}

It is obvious that $W_m$ in \equref{eq_Wm} depends on not only the frequency $\omega$ (given by the dynamic frequency shift), but also the phase angle $\theta$ (given by the modulation-demodulation). If considering the $\theta$-induced dynamics only here and ignoring the previously-discussed $\omega$-induced dynamics by setting $g_{mn} = G_{mn}(j\omega_0)$, in this case, we can re-write $W_m$ in \equref{eq_Wm} as
\begin{equation}    \label{eq_WmSim}
\begin{aligned}
W_m  & = \text{Re} \left( e^{-j\mu_m} \sum\nolimits_n { G_{mn} A_m A_n e^{{-j(\theta_m-\theta_n)}}} \right)
\\
 & = \sum\nolimits_n \Gamma_{mn} \sin(\theta_m - \theta_n + \gamma_{mn})
 \end{aligned}
\end{equation}
with
\begin{equation}
\begin{aligned}
& \Gamma_{mn}=|G_{mn} A_m A_n|
\\
& \gamma_{mn}= {\pi}/{2} +\mu_m -\arg G_{mn}
\end{aligned}
\end{equation}
$\Gamma_{mn}$ is the generalized synchronization coefficient, and $\gamma_{mn}$ is the offset angle. Equation \equref{eq_WmSim} shows interesting properties. $\Gamma_{mn}$ is symmetric ($\Gamma_{mn} = \Gamma_{nm}$) due to the reciprocity of electrical networks. The offset angle $\gamma_{mn}$ is dependent on network topology as well as the projection angle $\mu_m$ for different apparatuses. In the following two conditions, $\gamma_{mn}$ is approximately zero for $m \neq n$: (i) synchronization of grid-forming apparatuses via inductive transmission lines; and (ii) synchronization of grid-following inverters via shunt resistances (passive loads). In such cases, the synchronization equation (\ref{eq_sync}) is reduced to a second-order Kuramoto model which has a wide stability region \cite{peng2015synchronization}. This implies that grid-following converters could also have guaranteed stability that is similar to grid-forming apparatuses.

For cases where grid-forming and grid-following apparatuses co-exist, or where the power network is not purely inductive or resistive, $\gamma_{mn} \neq 0$, and the whole-system dynamic behaviour is more complicated. We can use linearization to evaluate the stability subject to small disturbances. Define $K_{mn} \triangleq \partial W_m / \partial \theta_n$ and $[K_H] \triangleq [H]^{-1} [K]$, $[K]$ is the matrix of $K_{mn}$, and $[H]$ is the diagonal matrix of $H_m$. $[K_H]$ is an extension of the synchronizing power coefficients in conventional power systems so that both voltage and current nodes can be considered \cite{kundur1994power}. The eigenvalue of $[K_H]$ determines the synchronizing capability of (\ref{eq_sync}), and the eigenvector determines the modal participation \cite{kundur1994power}. We define the critical eigenvalue $\xi_\text{c}$ as the non-zero eigenvalue of $[K_H]$ that has the minimum real part. The system is small-disturbance stable if $\xi_\text{c}$ has a positive real part, as derived in \appendixref{Appendix:SmallSignalStability}.

\section{Test of The IEEE 68-Bus Power System}

We verified our theoretic findings on the modified IEEE 68-bus system, and the test results are summarised in \figref{fig_K} (small-disturbance analysis) and \figref{fig_K_SimResult} (EMT simulations). All parameters, scripts, models are available online \cite{FuturePowerNetworks}. We tested five cases with different proportions of grid-following converters, subject to small and large disturbances. \figref{fig_K}(a)-(b) and \figref{fig_K_SimResult}(a)-(b) contain 100\% grid-following converters with passive and active loads, and are stable under both small and large disturbances. We gradually replaced the converters by synchronous generators in \figref{fig_K}(c)-(e), and found more complicated stability patterns. These agree well with the prediction of our theory. The critical eigenvalue $\xi_\text{c}$ of $[K_H]$ provides accurate indication of small-disturbance stability in all cases. We also display in \figref{fig_K}(c) the participation of each node in the critical eigenvalue, to show how $[K_H]$ helps to trace the origins of instability.

It is rather surprising to see that a power system with 100\% grid-following converters is rather stable and re-synchronises rather fast after faults, as shown in \figref{fig_K}(a)-(b) and \figref{fig_K_SimResult}(a)-(b). It is also worth highlighting that the voltage at each bus node is also nearby 1 pu at steady-state, by properly setting the current reference of each grid-following inverter based on the power flow analysis. It is even more surprising to see that adding one synchronous generator to the all-grid-following-converter network destabilises the system, as shown in \figref{fig_K}(c), which contradicts the conventional observation that synchronous generators are always helpful for grid stability. Because in this case, the current nodes dominate the power systems and the only one voltage node follows current nodes, which means worse stability than conventional case where current nodes follow voltage nodes, i.e., asymmetry of synchronization stability between voltage nodes and current nodes \cite{li2022revisiting}. This also raises the important issue about the placement of grid-forming apparatuses in a converter dominated grid. We use two techniques to guide this placement: participation analysis for the critical eigenvalue for small-disturbance stability, and observation of loss of synchronization via time-domain simulation for large-disturbance analysis. These techniques are effective as verified in \figref{fig_K}(c)-(e) and \figref{fig_K_SimResult}(c)-(d), showing succeeding improvements of stability when synchronous generators are placed at the nodes identified most influential (14 and 7). 


\section{Conclusions} \label{Section:Conclusions}

The power-communication isomorphism theory reveals the intrinsic analogy of power systems and communication systems. This analogy (isomorphism) can be used to interpret the synchronization stability of power systems from a communication perspective. The power-based and signal-based synchronization schemes are unified into a common principle. The dynamic channel gain reveals the role of EMT dynamics in phasor-domain analysis. The channel bandwidth influences the speed of power transfer and angle synchronization. The channel-frequency dependency (dynamic frequency shift) could worsen the system damping. Additionally, the synchronization capability of multi-bus power systems is discussed. We demonstrate that a 100\% grid-following-converter grid can be well stabilised (in both small- and large-signal sense) without the existence of any voltage source. We also illustrate that adding only one grid-forming apparatuses into a 100\% grid-following-converter grid could destabilize the system, but keep adding more grid-forming apparatuses can then enhance the system stability. The findings are verified on the modified IEEE 68-bus test system.



\appendices

\section{Derivation of Channel Bandwidth}   \label{Appendix:ChannelBandwidth}

The channel bandwidth equations (\ref{eq_pert})-(\ref{eq_filter}) are obtained as follows. Applying angle perturbations at both ends of the channel $g_{mn}$, the corresponding complex power perturbation is 
\begin{equation}    \label{eq_Delta_Smn}
{\Delta}S_{mn} = S_{mn0}g_{mn0}(\Delta {\vartheta_m}^{\!\!\!\!\!*} + \Delta{\vartheta_n} + g_{mn0}^{-1} \Delta g_{mn}).
\end{equation}
Linearising the channel gain equation (\ref{eq_gain_diff}) yields
\begin{equation}
d {\Delta g}/dt = (p - \varpi_0) \Delta g - g_0 \Delta \varpi
\end{equation}
from which we get the transfer function from $\Delta \varpi$ to $\Delta g$
\begin{equation}
{\Delta g}(s) = -\frac{g_0}{s + \varpi_0 -p} \Delta \varpi(s).
\end{equation}
Therefore,
\begin{equation}
g_{mn0}^{-1} \Delta g_{mn}(s) = \frac{-\Delta \varpi_n(s)}{s + \varpi_0 - p} = \frac{-s \Delta \vartheta_n(s)}{s + \varpi_0 - p}.
\end{equation}
Putting this into (\ref{eq_Delta_Smn}) and noting that $\varpi_0 = j\omega_0$ (since the signal amplitude is constant in steady-state), we get the equations (\ref{eq_pert})-(\ref{eq_filter}). 

\section{Hints on Calculating Damping and Synchronizing Coefficients} \label{Appendix:CalculationHint}

The complex power from node $n$ to node $m$ is
\begin{equation}
    S_{mn} = g_{mn} \underbrace{e^{\vartheta_n}e^{\vartheta_m^*}}_{\hat{S}_{mn}} \approx \underbrace{G_{mn}(\varpi_n)}_{f(\varpi)} \underbrace{e^{\vartheta_n}e^{\vartheta_m^*}}_{f(\vartheta)}
\end{equation}
where $f(\varpi)$ contributes to the damping power, and $f(\vartheta)$ contributes to the synchronizing power. The total complex power at node $m$ can be represented as 
\begin{equation}
    S_m = \sum\nolimits_n S_{mn} = A_m^2 G_{mm} + e^{\vartheta_m^*} \sum\nolimits_{n\neq m} G_{mn} e^{\vartheta_n}
\end{equation}

For calculating damping coefficient, $\partial S_{mn}/ \partial \varpi_n$ can be represented as
\begin{equation}
    \frac{\partial S_{mn}}{\partial \varpi_n} 
    = \frac{\partial G_{mn}(\varpi_n)}{\partial \varpi_n} e^{\vartheta_n}e^{\vartheta_m^*} 
    = G_{mn}^\prime e^{\vartheta_n}e^{\vartheta_m^*} 
\end{equation}
for both $n \neq m$ and $n = m$. Based on it, we can further get
\begin{equation}
    \frac{\partial S_{m}}{\partial \omega_n} = \frac{\partial S_{mn}}{\partial \omega_n} = jG_{mn}^\prime e^{\vartheta_n}e^{\vartheta_m^*} 
\end{equation}
for both $n \neq m$ and $n = m$.

For calculating the synchronizing coefficient, $\partial S_{mn}/ \partial \vartheta_n$ can be represented as
\begin{equation} \label{Eq:SynchronizingCoefficient}
\begin{aligned}
    & \frac{\partial S_{mn}}{\partial \vartheta_n} = G_{mn}e^{\vartheta_n}e^{\vartheta_m^*} = S_{mn0}, ~\text{if}~ n \neq m
    \\
    & \frac{\partial S_{mm}}{\partial \vartheta_m} = 0,~\text{if}~ n = m
\end{aligned}
\end{equation}
Based on it, we can further get
\begin{equation}
    \begin{aligned}
    & \frac{\partial S_{m}}{\partial \theta_n} = \frac{\partial S_{mn}}{\partial \theta_n} = j G_{mn}e^{\vartheta_n}e^{\vartheta_m^*}, ~\text{if}~ n \neq m
    \\
    & \frac{\partial S_{m}}{\partial \theta_m} = -j \sum\nolimits_{n \neq m} G_{mn} e^{\vartheta_n}e^{\vartheta_m^*},~\text{if}~ n = m
\end{aligned}
\end{equation}

It is also worth mentioning that 
\begin{equation}
    \frac{\partial W}{\partial \omega} 
    = \frac{\partial \text{Re}(e^{-j\mu}S)}{\partial \omega} = \text{Re}\Big( e^{-j\mu}\frac{\partial S}{\partial \omega} \Big)
\end{equation}
similarly for $\partial W / \partial \theta$. 

\section{Small Disturbance Stability of Synchronization} \label{Appendix:SmallSignalStability}
Here, we show why the  small-disturbance stability of the whole system is determined by the eigenvalues of $[K_H]$. Linearizing (\ref{eq_sync}) yields
\begin{equation}
\label{eq_H_omega}
[\Delta \ddot{\theta}] = -[H]^{-1}[D][\Delta \dot{\theta}] - [K_H][\Delta \theta]
\end{equation}
The generalized inertia and damping are usually proportional \cite{kundur1994power}, so we have $[H]^{-1}[D] = \sigma [I]$ where $[I]$ is a unit matrix and $\sigma = D_m/H_m$. $[K_H]$ can be diagonalized by $[K_H] = [\Phi] [\Xi] [\Phi]^{-1}$ where $[\Xi]$ is a diagonal matrix containing the eigenvalues of $[K_H]$, and $[\Phi]$ contains the corresponding eigenvectors. Define the coordinate transformation $[\Phi]^{-1} [\Delta \theta] = [z]$, and transform (\ref{eq_H_omega}) to the $z$ coordinate
\begin{equation}
\label{eq_z_ss}
\ddot{[z]} = -\sigma \dot{[z]} - [\Xi][z].
\end{equation}
Since $[\Xi]$ is diagonal, (\ref{eq_z_ss}) reduces to a series of decoupled second order systems
\begin{equation}
\label{eq_z_sec}
\ddot{z}_m = -\sigma \dot{z}_m - \xi_m z_m
\end{equation}
where $m \in \{1,2, \cdots ,N\}$ and $\xi_m$ is the $m$-th eigenvalue of $K_H$. The system (\ref{eq_z_sec}) is stable if an only if its characteristic equation $s^2 + \sigma s + \xi_m = 0$ only has solutions in the left open half complex plane. We traverse $s$ in the unstable right half plane to get the forbidden region of $\xi_m$, and the stable region is its complement
\begin{equation}
\label{eq_xi_reg}
\text{Stable Region: } 
\{\xi_m \ | \  \text{Re} \,\xi_m > \sigma^{-2} (\text{Im} \, \xi_m)^2\}.
\end{equation}
If sufficient damping is provided in synchronisation control, $\sigma$ is large enough to render $\sigma^{-2} (\text{Im} \, \xi_m)^2$ very small, and the stability region is approximated by  
\begin{equation}
\text{Stable Region: } 
\{\xi_m \ | \  \text{Re} \,\xi_m > 0\}.
\end{equation}
All $\xi_m$ must be within the stable region to ensure the synchronisation stability of the power system, with the only exception being the one that equals zero, which represents the collective spinning of the entire power system. Therefore we define the critical eigenvalue $\xi_\text{c}$ as the non-zero $\xi_m$ (eigenvalue of $[K_H]$) that has the minimum real part.

\ifCLASSOPTIONcaptionsoff
  \newpage
\fi

\bibliographystyle{IEEEtran}
\bibliography{Paper}

\begin{thebibliography}{10}
\providecommand{\url}[1]{#1}
\csname url@samestyle\endcsname
\providecommand{\newblock}{\relax}
\providecommand{\bibinfo}[2]{#2}
\providecommand{\BIBentrySTDinterwordspacing}{\spaceskip=0pt\relax}
\providecommand{\BIBentryALTinterwordstretchfactor}{4}
\providecommand{\BIBentryALTinterwordspacing}{\spaceskip=\fontdimen2\font plus
\BIBentryALTinterwordstretchfactor\fontdimen3\font minus
  \fontdimen4\font\relax}
\providecommand{\BIBforeignlanguage}[2]{{%
\expandafter\ifx\csname l@#1\endcsname\relax
\typeout{** WARNING: IEEEtran.bst: No hyphenation pattern has been}%
\typeout{** loaded for the language `#1'. Using the pattern for}%
\typeout{** the default language instead.}%
\else
\language=\csname l@#1\endcsname
\fi
#2}}
\providecommand{\BIBdecl}{\relax}
\BIBdecl

\bibitem{bialek2020does}
J.~Bialek, ``What does the {GB} power outage on 9 august 2019 tell us about the
  current state of decarbonised power systems?'' \emph{Energy Policy}, vol.
  146, p. 111821, 2020.

\bibitem{kundur1994power}
P.~Kundur, \emph{Power system stability and control}.\hskip 1em plus 0.5em
  minus 0.4em\relax McGraw-hill New York, 1994, vol.~7.

\bibitem{kundur2004definition}
P.~Kundur \emph{et~al.}, ``Definition and classification of power system
  stability ieee/cigre joint task force on stability terms and definitions,''
  \emph{IEEE Trans. Power Syst.}, vol.~19, no.~3, pp. 1387--1401, 2004.

\bibitem{rocabert2012control}
J.~Rocabert, A.~Luna, F.~Blaabjerg, and P.~Rodríguez, ``Control of power
  converters in {AC} microgrids,'' \emph{IEEE Trans. Power Electron.}, vol.~27,
  no.~11, pp. 4734--4749, Nov. 2012.

\bibitem{rosso2021grid}
R.~Rosso, X.~Wang, M.~Liserre, X.~Lu, and S.~Engelken, ``Grid-forming
  converters: Control approaches, grid-synchronization, and future trends—a
  review,'' \emph{IEEE Open Journal of Industry Applications}, vol.~2, pp.
  93--109, 2021.

\bibitem{matevosyan2019grid}
J.~Matevosyan, B.~Badrzadeh, T.~Prevost, E.~Quitmann, D.~Ramasubramanian,
  H.~Urdal, S.~Achilles, J.~MacDowell, S.~H. Huang, V.~Vital \emph{et~al.},
  ``Grid-forming inverters: Are they the key for high renewable penetration?''
  \emph{IEEE Power and Energy magazine}, vol.~17, no.~6, pp. 89--98, 2019.

\bibitem{li2022revisiting}
Y.~Li, Y.~Gu, and T.~C. Green, ``Revisiting grid-forming and grid-following
  inverters: A duality theory,'' \emph{IEEE Transactions on Power Systems},
  2022.

\bibitem{fan2020identifying}
L.~Fan, Z.~Miao, P.~Koralewicz, S.~Shah, and V.~Gevorgian, ``Identifying
  dq-domain admittance models of a 2.3-mva commercial grid-following inverter
  via frequency-domain and time-domain data,'' \emph{IEEE Transactions on
  Energy Conversion}, vol.~36, no.~3, pp. 2463--2472, 2020.

\bibitem{wen2015analysis}
B.~Wen, D.~Boroyevich, R.~Burgos, P.~Mattavelli, and Z.~Shen, ``Analysis of dq
  small-signal impedance of grid-tied inverters,'' \emph{IEEE Transactions on
  Power Electronics}, vol.~31, no.~1, pp. 675--687, 2015.

\bibitem{wang2020grid}
X.~Wang, M.~G. Taul, H.~Wu, Y.~Liao, F.~Blaabjerg, and L.~Harnefors,
  ``Grid-synchronization stability of converter-based resources—an
  overview,'' \emph{IEEE Open Journal of Industry Applications}, vol.~1, pp.
  115--134, 2020.

\bibitem{sof2018system}
{System Operability Framework}, ``{Performance of Phase-locked Loop based
  converters},'' National Grid, Tech. Rep., 2018.

\bibitem{harnefors2007modeling}
L.~Harnefors, ``Modeling of three-phase dynamic systems using complex transfer
  functions and transfer matrices,'' \emph{IEEE Trans. Ind. Electron.},
  vol.~54, no.~4, pp. 2239--2248, Aug. 2007.

\bibitem{briz2000analysis}
F.~Briz, M.~W. Degner, and R.~D. Lorenz, ``Analysis and design of current
  regulators using complex vectors,'' \emph{IEEE Transactions on Industry
  Applications}, vol.~36, no.~3, pp. 817--825, 2000.

\bibitem{martin2004complex}
K.~W. Martin, ``Complex signal processing is not complex,'' \emph{IEEE
  Transactions on Circuits and Systems I: Regular Papers}, vol.~51, no.~9, pp.
  1823--1836, 2004.

\bibitem{milano2021complex}
F.~Milano, ``Complex frequency,'' \emph{IEEE Transactions on Power Systems},
  vol.~37, no.~2, pp. 1230--1240, 2021.

\bibitem{pattabiraman2018comparison}
D.~Pattabiraman, R.~Lasseter, and T.~Jahns, ``{Comparison of grid following and
  grid forming control for a high inverter penetration power system},'' in
  \emph{2018 IEEE Power \& Energy Society General Meeting (PESGM)}.\hskip 1em
  plus 0.5em minus 0.4em\relax IEEE, 2018, pp. 1--5.

\bibitem{czarnecki2004some}
L.~S. Czarnecki, ``On some misinterpretations of the instantaneous reactive
  power pq theory,'' \emph{IEEE transactions on power electronics}, vol.~19,
  no.~3, pp. 828--836, 2004.

\bibitem{zhang2009power}
L.~Zhang, L.~Harnefors, and H.-P. Nee, ``Power-synchronization control of
  grid-connected voltage-source converters,'' \emph{IEEE Transactions on Power
  systems}, vol.~25, no.~2, pp. 809--820, 2009.

\bibitem{2014virtual}
S.~{D'Arco} and J.~A. {Suul}, ``{Equivalence of Virtual Synchronous Machines
  and Frequency-Droops for Converter-Based MicroGrids},'' \emph{IEEE
  Transactions on Smart Grid}, vol.~5, no.~1, pp. 394--395, Jan 2014.

\bibitem{chung2000phase}
S.-K. Chung, ``A phase tracking system for three phase utility interface
  inverters,'' \emph{IEEE Trans. Power Electron.}, vol.~15, no.~3, pp.
  431--438, May 2000.

\bibitem{chen2004electrical}
W.~K. Chen, \emph{The electrical engineering handbook}.\hskip 1em plus 0.5em
  minus 0.4em\relax Elsevier, 2004.

\bibitem{gu2020impedance}
Y.~Gu, Y.~Li, Y.~Zhu, and T.~C. Green, ``Impedance-based whole-system modeling
  for a composite grid via embedding of frame dynamics,'' \emph{IEEE
  Transactions on Power Systems}, vol.~36, no.~1, pp. 336--345, 2020.

\bibitem{oppeitheim1983signals}
A.~Oppeitheim, A.~S. Willsky, and I.~Young, ``Signals and systems,''
  \emph{Prentice-Hall, Englewood Cliffs, New Jersey}, vol.~19, pp. 146--153,
  1983.

\bibitem{li2022mapping}
Y.~Li, Y.~Gu, and T.~Green, ``Mapping of dynamics between mechanical and
  electrical ports in {SG}-{IBR} composite grids,'' \emph{IEEE Transactions on
  Power Systems}, 2022.

\bibitem{peng2015synchronization}
J.~Peng, ``Synchronization in the second-order kuramoto model,'' 2015.

\bibitem{FuturePowerNetworks}
\BIBentryALTinterwordspacing
``{Future Power Networks}.'' [Online]. Available:
  \url{https://github.com/Future-Power-Networks/Publications}
\BIBentrySTDinterwordspacing

\end{thebibliography}

\end{document}